\documentclass[aps,prl,twocolumn,groupedaddress,superscriptaddress]{revtex4-1}
\usepackage{graphicx}
\usepackage{amssymb}
\usepackage{epstopdf}
\usepackage{amsmath}
\usepackage{longtable}
\usepackage{amsmath}
\usepackage{amsfonts}
\usepackage[colorlinks=true,citecolor=blue,linkcolor=red]{hyperref}
\usepackage{graphicx}
\usepackage{bbold}     
\usepackage[dvipsnames]{xcolor}

\usepackage{array,booktabs,ragged2e}
\usepackage{soul}      
\usepackage{cancel}      
\usepackage{tabularx}     
\usepackage{multirow}
\usepackage{hhline}
\usepackage{titlesec}

\begin{document}

\title{ Extremely large magnetoresistance in the ``ordinary'' metal ReO$_3$}

\author{Qin Chen}
\affiliation{Department of Physics, Zhejiang University, Hangzhou 310027, China}
\author{Zhefeng Lou}
\affiliation{Department of Physics, Zhejiang University, Hangzhou 310027, China}
\author{ShengNan Zhang}
\affiliation{Institute of Physics, \'{E}cole Polytechnique F\'{e}d\'{e}rale de Lausanne (EPFL), CH-1015 Lausanne, Switzerland}
\affiliation{National Centre for Computational Design and Discovery of Novel Materials MARVEL, \'{E}cole Polytechnique F\'{e}d\'{e}rale de Lausanne (EPFL), CH-1015 Lausanne, Switzerland}
\author{Yuxing Zhou}
\affiliation{Department of Physics, Zhejiang University, Hangzhou 310027, China}
\author{Binjie Xu}
\affiliation{Department of Physics, Zhejiang University, Hangzhou 310027, China}
\author{Huancheng Chen}
\affiliation{Department of Physics, Zhejiang University, Hangzhou 310027, China}
\author{Shuijin Chen}
\affiliation{Department of Physics, Zhejiang University, Hangzhou 310027, China}
\author{Jianhua Du}
\affiliation{Department of Applied Physics, China Jiliang University, Hangzhou $310018$, China}
\author{Hangdong Wang}
\affiliation{Department of Physics, Hangzhou Normal University, Hangzhou 310036, China}
\author{Jinhu Yang}
\affiliation{Department of Physics, Hangzhou Normal University, Hangzhou 310036, China}
\author{QuanSheng Wu}
\affiliation{Institute of Physics, \'{E}cole Polytechnique F\'{e}d\'{e}rale de Lausanne (EPFL), CH-1015 Lausanne, Switzerland}
\affiliation{National Centre for Computational Design and Discovery of Novel Materials MARVEL, \'{E}cole Polytechnique F\'{e}d\'{e}rale de Lausanne (EPFL), CH-1015 Lausanne, Switzerland}
\author{Oleg V. Yazyev}
\affiliation{Institute of Physics, \'{E}cole Polytechnique F\'{e}d\'{e}rale de Lausanne (EPFL), CH-1015 Lausanne, Switzerland}
\affiliation{National Centre for Computational Design and Discovery of Novel Materials MARVEL, \'{E}cole Polytechnique F\'{e}d\'{e}rale de Lausanne (EPFL), CH-1015 Lausanne, Switzerland}
\author{Minghu Fang}\email{Corresponding author: mhfang@zju.edu.cn}
\affiliation{Department of Physics, Zhejiang University, Hangzhou 310027, China}
\affiliation{Collaborative Innovation Center of Advanced Microstructure, Nanjing University, Nanjing 210093, China}
\date{\today}
\begin{abstract}

The extremely large magnetoresistance (XMR) observed in many topologically nontrivial and trivial semimetals has attracted much attention in relation to its underlying physical mechanism. In this paper, by combining the band structure and Fermi surface (FS) calculations with the Hall resistivity and de Haas-Van Alphen (dHvA) oscillation measurements, we studied the anisotropy of magnetoresistance (MR) of ReO$_3$ with a simple cubic structure, an ``ordinary" nonmagnetic metal considered previously. We found that ReO$_3$ exhibits almost all the characteristics of XMR semimetals: the nearly quadratic field dependence of MR, a field-induced upturn in resistivity followed by a plateau at low temperatures, high mobilities of charge carriers. It was found that for magnetic field \emph{H} applied along the \emph{c} axis, the MR exhibits an unsaturated \emph{H}$^{1.75}$ dependence, which was argued to arise from the complete carrier compensation supported by the Hall resistivity measurements. For \emph{H} applied along the direction of 15$^\circ$ relative to the \emph{c} axis, an unsaturated \emph{H}$^{1.90}$ dependence of MR up to 9.43~$\times$~$10^3$$\%$ at 10~K and 9~T was observed, which was explained by the existence of electron open orbits extending along the $k_{x}$ direction. Two mechanisms responsible for XMR observed usually in the semimetals occur also in the simple metal ReO$_3$ due to its peculiar FS (two closed electron pockets and one open electron pocket), once again indicating that the details of FS topology are a key factor for the observed XMR in materials.

\end{abstract}
\maketitle

\section{I. INTRODUCTION}
The fundamental and applied research on the magnetoresistance (MR) attracted a lot of attention in the past 30 years, due to its applications in magnetic devices for data storage \cite{moritomo1996giant,lenz1990review,daughton1999gmr}, magnetic valves \cite{wolf2001spintronics}, magnetic sensors or magnetic switches \cite{lenz1990review,jankowski2011hall}. The recent discovery of the extremely large magnetoresistance (XMR) up to $10^6$$\%$ at low temperatures in numerous compounds motivates further research into MR. The list of XMR materials includes both the topologically nontrivial compounds, such as the Dirac semimetals Na$_3$Bi \cite{liu2014discovery} and Cd$_3$As$_2$ \cite{li2016negative}, Weyl semimetals of TaAs family \cite{sun2015topological}, WTe$_2$ \cite{thoutam2015temperature}, $\beta$-WP$_2$ \cite{wang2017large}, elemental Ga \cite{chen2018large}, MoO$_2$ \cite{PhysRevB.102.165133} and VAs$_2$ \cite{chen2021magnetoresistance} as well as topologically trivial semimetals, such as elemental Bi \cite{PhysRevLett.9.62}, PdCoO$_2$ \cite{takatsu2013extremely}, PtSn$_4$ \cite{PhysRevB.97.205132}, transition metal dipnictides TPn$_2$ (T = Ta and Nb, Pn = P, As and Sb) \cite{PhysRevB.93.195119,wu2016giant,PhysRevB.93.195106,PhysRevB.94.041103,PhysRevB.94.121115,PhysRevB.93.184405,luo2016anomalous}, $\alpha$-WP$_2$ \cite{PhysRevB.97.245101}, rock salt rare earth compound LaBi/Sb \cite{PhysRevB.93.241106,PhysRevLett.117.127204}, SiP$_2$ \cite{zhou2020linear} and many others. Although the family of materials showing XMR is expanding, no consistent explanation for the XMR mechanism has been developed so far. Nontrivial band topology inducing linear band dispersion was believed to be responsible for the linear field dependent MR in Cd$_3$As$_2$ \cite{PhysRevB.92.081306}. Classical charge-carrier compensation scenario was invoked to explain the non-saturating quadratic MR in WTe$_2$ \cite{thoutam2015temperature}. Open-orbit trajectories of charge carriers as a result of non-closed Fermi furface (FS) was employed to illustrate the XMR behavior in PdCoO$_2$ \cite{takatsu2013extremely}. Recently, Zhang~\emph{et al}. \cite{PhysRevB.99.035142} studied the transverse MR by combining the FS calculations with the Boltzmann transport theory and the relaxation time approximation, finding that the details of FS topology plays an important role in both the field dependence of MR and its anisotropy.

ReO$_3$ crystallizes in a simple cubic structure with space group \emph{P}m$\bar{3}$m (No.~221). As an ``ordinary" nonmagnetic metallic oxide, its calculated band structure and FS \cite{PhysRev.181.987} compare well with the de Haas-Van Alphen (dHvA) oscillation \cite{marcus1968measurement} and optical spectroscopy \cite{PhysRev.165.765} measurements. The ``compressibility-collapse" transition occurring in ReO$_3$ also attracted considerable  attention \cite{PhysRevB.24.692,PhysRevLett.42.1485,razavi1978pressure}, especially for the micro-structure change at this transition by using nuclear magnetic resonance (NMR) measurements \cite{PhysRevB.20.4746}, and FS change by using the symmetry analysis \cite{PhysRevB.24.692}. Meanwhile, as a comparison with the copper oxide superconductors exhibiting anomalous resistivity in the normal state, the temperature dependence of the longitudinal resistivity and Hall resistivity of ReO$_3$ was also analyzed  \cite{PhysRevB.47.14434} by using the Bloch-Gr\"{u}neisen form as a strong electron-phonon coupling metal.

Usually, in most nonmagnetic metals MR is a relatively weak effect, characterized by a quadratic field dependence at low fields that saturates to a magnitude of a few percent at higher fields, totally different from that in semimetals. In this paper, we measured the longitudinal resistivity, $\rho_{xx}$(\emph{T},\emph{H}), Hall resistivity, $\rho_{xy}$(\emph{T},\emph{H}), and dHvA oscillations, as well as calculated the band structure and FS of ReO$_3$. We find a nearly quadratic field dependence of MR reaching a large value of 9.43~$\times$~10$^3$$\%$ at 10~K and 9~T, as well as a field-induced up-turn behavior of $\rho$$_{xx}$(\emph{T}), which are the common characteristics for many topologically nontrivial and trivial semimetals. For magnetic field \emph{H} applied along the \emph{c} axis, the MR exhibits a non-saturating \emph{H}$^{1.75}$ dependence, which we argue arises from the carrier compensation, evidenced by the Hall resistivity measurements. For magnetic field applied along other directions, a similar non-saturating MR dependence with \emph{H}$^{n}$ (\emph{n} = 1.68$-$1.90) was observed, which we conclude is due to the existence of electron open orbits extending along the $k_{x}$ direction. These results indicate that the details of FS topology play an important role in the anisotropy of MR.

\section{II. EXPERIMENTAL AND COMPUTATIONAL METHODS}
\begin{figure*}[htbp]
\centering
\includegraphics[width=17.5cm]{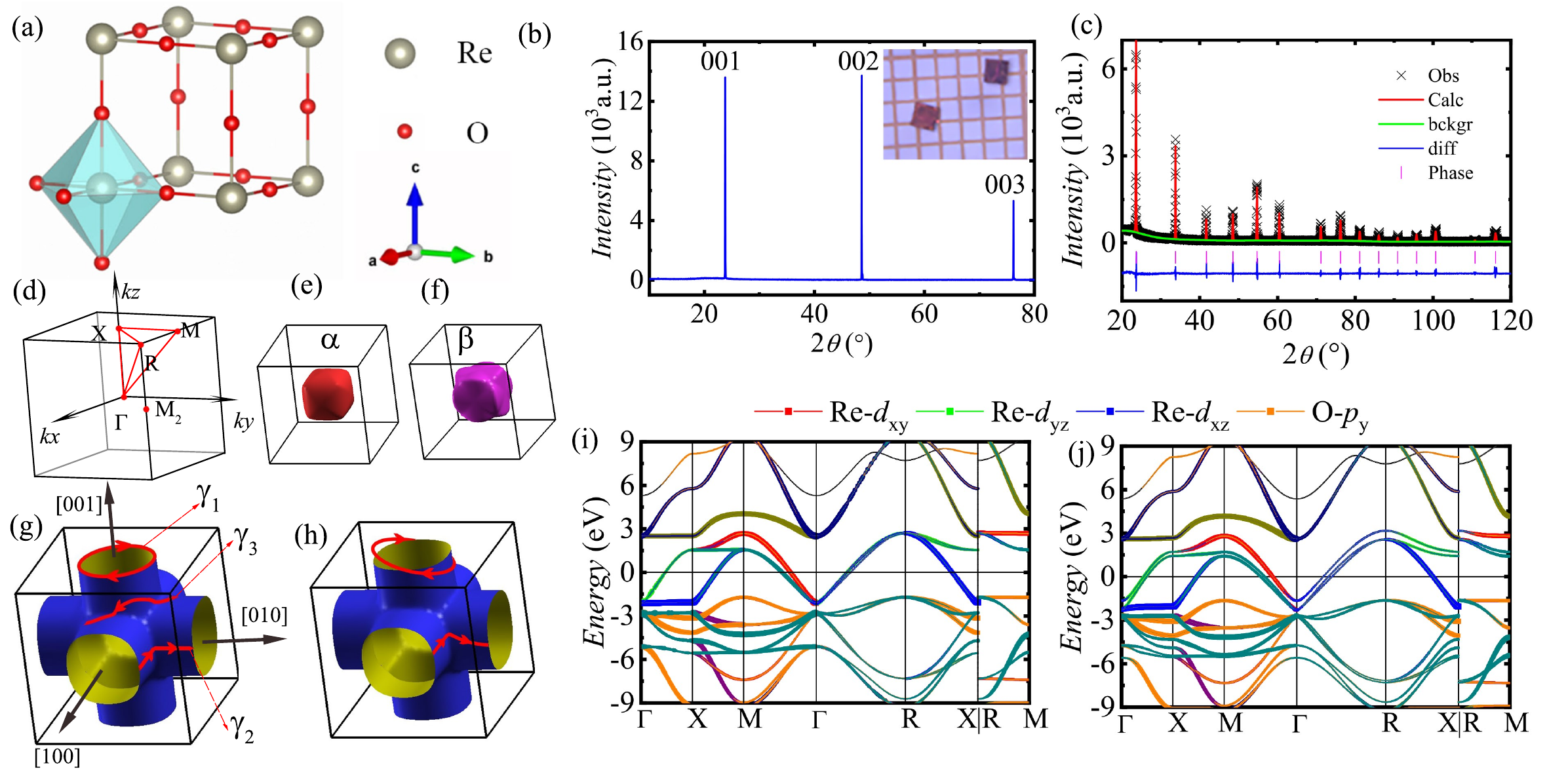}
\caption{(a) Crystal structure of cubic ReO$_3$. (b) XRD pattern of a ReO$_3$ single crystal. (c) XRD pattern of powder obtained by grinding ReO$_3$ crystals, the line shows its Rietveld refinement. (d) The Brillouin zone of ReO$_3$. (e), (f), (g) and (h) Three dimensional views of electron Fermi pockets. (i) and (j) Band structures of ReO$_3$ calculated without and with spin-orbit coupling.}
\label{fig1}
\end{figure*}

ReO$_3$ single crystals were grown by a chemical vapor transport method. Polycrystalline ReO$_3$ prepared previously was sealed in an evacuated quartz tube with 10 mg/cm$^3$ TeCl$_4$ as a transport agent, then heated for 2 weeks at 670~K, in a tube furnace with a gradient 30~K. Red crystals with typical dimensions 1.0~$\times$~1.0~$\times$~0.2~mm$^3$ and a (001) easy cleavage plane [see Fig.~\ref{fig1}(b)] were obtained at the cold end of the tube. The composition was confirmed to be Re:O~=~1:3 by using the energy dispersive x-ray spectrometer (EDXS). The crystal structure was determined using a powder x-ray diffractometer (XRD, Rigaku Gemini A Ultra) with samples produced by grinding pieces of crystals [see Fig.~\ref{fig1}(b)]. It was confirmed that ReO$_3$ crystallizes in a cubic structure (space group \emph{P}m$\bar{3}$m, No. 221). The lattice parameters \emph{a} =  \emph{b} = \emph{c} = 3.750(2) $\rm {\AA}$ were obtained by using the Rietveld refinement to XRD data (weighted profile factor R$_{wp}$ = 9.62$\%$, and the goodness-of-fit $\chi$$^2$ = 2.540), as shown in Fig.~\ref{fig1}(c). Electrical resistivity ($\rho$$_{xx}$), Hall resistivity ($\rho$$_{xy}$), and magnetization measurements were carried out by using a Quantum Design physical property measurement system (PPMS - 9 T) or Quantum Design magnetic property  measurement system (MPMS - 7 T).

The band structure calculations were performed using the Vienna \textit{ab} \textit{initio} simulation package (VASP) \cite{kresse199614251,kresse1999g} with the generalized gradient approximation (GGA) of Perdew, Burke and Ernzerhof (PBE) \cite{perdew1996generalized} for the exchange-correlation potential. A cutoff energy of 520 eV and a $13\times13\times13$ k-point mesh were used to perform the bulk calculations.
Magnetoresistance was calculated using the combination of the Boltzmann transport theory and the Fermi surface obtained from first principles \cite{zhang2019magnetoresistance}.
For this purpose we used the WannierTools~\cite{wu2018wanniertools} package, which
is based on the maximally localized Wannier function tight-binding model \cite{marzari1997, souza2001, marzari2012} constructed by using the Wannier90 \cite{mostofi2014updated} package.

Within the relaxation time approximation, the band-wise conductivity tensor $\bold{\sigma}$ is calculated by solving the Boltzmann equation in presence of an applied magnetic field as \cite{Neilbook,wu2018wanniertools,zhang2019magnetoresistance},
\begin{equation}
\sigma^{(n)}_{ij}(\bold{B})=\frac{ e^2}{4 \pi^3} \int d\bold{k} \tau_n \bold{v}_n(\bold{k})  \bold{\bar{v}}_n(\bold{k})
\left(- \frac{\partial f}{\partial \varepsilon} \right)_{\varepsilon=\varepsilon_n(\bold{k})},
\label{eqn-sigmaij}
\end{equation}
where $e$ is the electron charge, $n$ is the band index, $\tau_n$ is the relaxation time of $n$th band that is assumed to be independent on the wavevector $\bold{k}$, $f$ is the Fermi-Dirac distribution, $\bold{v}_n(\bold{k})$ is the velocity defined by the gradient of band energy
\begin{equation}
\bold{v}_n(\bold{k})=\frac{1}{\hbar} \nabla_{\bold{k}} \varepsilon_n(\bold{k}),
\label{eqn-velocity}
\end{equation}
and $\bar{\bold{v}}_n(\bold{k})$ is the weighted average of velocity over the past history of the charge carrier.
\begin{equation}
\bar{\bold{v}}_n(\bold{k}) = \int^0_{-\infty} \frac{dt}{\tau_n} e^{\frac{t}{\tau_n}} \bold{v}_n(\bold{k}(t)),
\label{eqn-aver_velo}
\end{equation}
The orbital motion of charge carriers in applied magnetic field causes the time evolution of $\bold{k}_n(t)$, written as,
\begin{equation}
\frac{d \bold{k}_n(t)}{dt} = - \frac{e}{\hbar} \bold{v}_n(\bold{k}(t)) \times \bold{B}.
\label{eqn-evol_k}
\end{equation}
with $ \bold{k}_n(0)=\bold{k}$.
The total conductivity is the sum of band-wise conductivities, \emph{i}.\emph{e}. $\sigma_{ij} = \sum_n \sigma^{(n)}_{ij} $, which is then inverted to obtain the resistivity tensor $\hat{\rho} = \hat{\sigma}^{-1}$.

\section{III. RESULTS AND DISCUSSION}
As a starting point, we discuss the results of our electronic band structure and FS calculations. Figures~\ref{fig1}(i) and  \ref{fig1}(j) show the bands without and with considering spin-orbit coupling (SOC), respectively. It is clear the states near the Fermi level (E$_F$) are composed of \emph{d} orbitals of Re atoms, while \emph{p} orbitals of O atoms are located at $-$2.5~eV relative to E$_F$, with the SOC resulting only in the separation of the bands near E$_F$. The three pockets of FS of ReO$_3$ are shown in Figs.~\ref{fig1}(e)$-$~\ref{fig1}(h), respectively, corresponding to the three electron-like surfaces centered at the $\Gamma$ point. The $\alpha$ and $\beta$ pockets are closed, while the $\gamma$ pockets is open along the [100] direction. The $\alpha$ pocket is rather circular in the (100) planes and slightly squared off in the (110) planes, the reverse is true for the $\beta$ pocket. The open $\gamma$ pocket consists of three intersecting cylinders. As discussed in Ref.~\cite{PhysRev.181.987,PhysRevB.24.692}, when magnetic field is applied along [001], two closed extremal orbits exist on the $\gamma$ pocket, as shown in Fig. \ref{fig1}(g). The electron-like orbit labelled $\gamma_1$ occurs on the arms of the cylinder and centered at the X point. The hole-like orbit labelled $\gamma_2$, closed in the extended zone scheme and centered at the M$_2$ point. When magnetic field is applied along the [111] direction, another extremal open orbit $\gamma_3$ exists on the $\Gamma$ pocket [see Fig. \ref{fig1}(g)]. Because of the nearly degenerate bands along the $\Gamma$-R direction [see Fig.~\ref{fig1}(d)], the $\beta$ and $\gamma$ pockets nearly touch along the [111] direction. The $\alpha$, $\beta$ and $\gamma$ pockets contain 0.093, 0.171 and 0.736 electrons per Re atom, respectively.

\begin{figure}[!htbp]
\centering
\includegraphics[width=8.6cm]{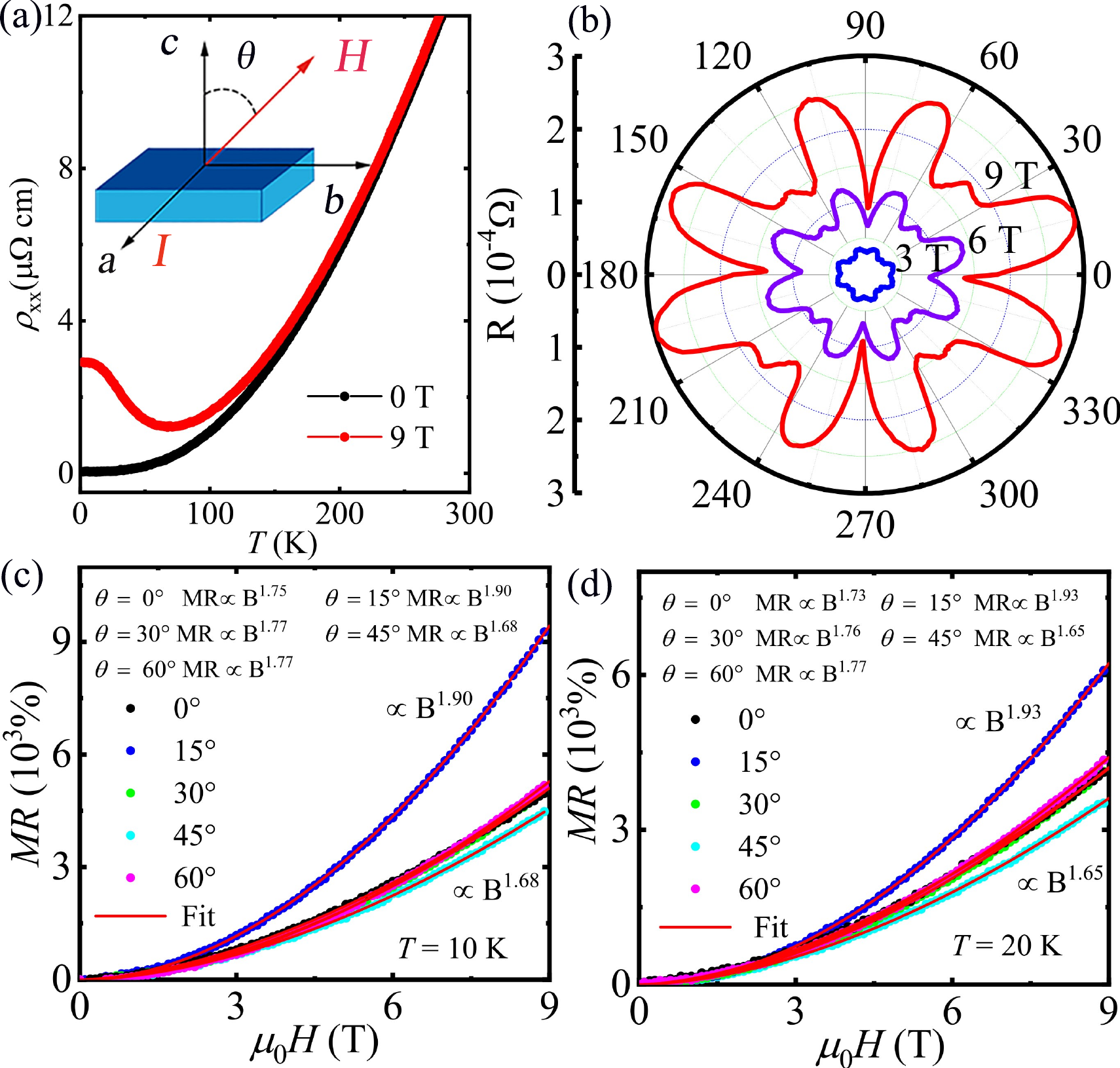}
\caption{(a) Temperature dependence of longitudinal resistivity, $\rho_{xx}$(\emph{T}), measured at 0~T and 9~T for a ReO$_3$ crystal. Schematic diagram of resistivity measurements: the current is applied along the \emph{a} axis and the field orientation $\theta$ is given in the \emph{b-c} plane (inset). (b) The angular plot of resistivity  measured at 2~K under various fields. (c) and (d) Magnetoresistance as a function of magnetic field for several chosen $H$ orientations measured at 10~K and 20~K, respectively.}\label{fig2}
\end{figure}

In order to study the role of the details of FS topology in the MR, considering the existence of both the closed and open pockets in ReO$_3$ mentioned above, we perfromed experimental measurements of the longitudinal resistivity and the MR anisotropy.  Figure~\ref{fig2}(a) shows the temperature dependence of the resistivity, $\rho_{xx}$(\emph{T}), measured at both $\mu_{0}$\emph{H} = 0~T and 9~T, respectively, with the current applied along the \emph{a} axis (\emph{I} $\parallel$ \emph{a}) and magnetic field applied along the \emph{c} axis. It is clear that $\rho$$_{xx}$ measured at $\mu_{0}$\emph{H} = 0~T decreases monotonically with decreasing temperature, \emph{i}.\emph{e}., exhibiting a typical  metallic behavior with $\rho$(2~K) = 0.02 $\mu$$\Omega$~cm and $\rho$(300~K) = 6.95 $\mu$$\Omega$~cm. This corresponds to residual resistivity ratio (RRR) of 348, indicating that our ReO$_3$ sample has a relatively high quality, which is consistent with that reported previously \cite{PhysRevB.10.2190,tanaka1976role,king1971low}. Interestingly,  $\rho_{xx}$(\emph{T}) measured at $\mu_{0}$\emph{H} = 9~T exhibits a  behavior similar to that  observed in both trivial topologically and nontrivial  semimetals \cite{takatsu2013extremely,mun2012magnetic,yuan2016large,huang2015observation,ali2014large,chen2016extremely} with a field-induced up-turn in resistivity followed by a plateau at low temperatures, indicating that a large MR indeed emerges in the nonmagnetic metal ReO$_3$.

Figure~\ref{fig2}(b) shows the angular resistance polar plot \emph{R}$_{xx}$ (\emph{H}, $\theta$) measured at 2~K in $\mu_0$\emph{H} = 3~T, 6~T, and 9~T with\emph{ I} along the \emph{a} axis and by rotating the magnetic field \emph{H } in the \emph{b}-\emph{c} plane [see the inset of Fig. \ref{fig2}(a)]. The \emph{R}($\theta$) at 2 K exhibits a nearly fourfold symmetry, \emph{i}.\emph{e}., \emph{R}($\theta$) = \emph{R}($\theta+\pi/2$), which is consistent with the cubic structure of ReO$_3$. An observed minor deviation is probably due to \emph{I} not being aligned exactly along the \emph{a} axis. The resistance grows quickly from a minimum at $\theta$ = 0$^\circ$ (\emph{H} $\parallel$ \emph{c} axis) to a maximum at $\theta$ = 15$^\circ$, and then decreases rapidly to another minimum at $\theta$ = 30$^\circ$, then increased the second maximum at $\theta$ = 45$^\circ$. As \emph{H} is rotated in the $\theta$ = 0 - 90$^\circ$ range, the resistance exhibits two maxima of one type ($\theta$ = 15$^\circ$ and 75$^\circ$) and a maximum of another type ($\theta$ = 45$^\circ$).

Then, we measured the field dependence of MR, with the conventional definition $\textit{MR} = \frac{\Delta\rho}{\rho(0)} = [\frac{\rho(H)-\rho(0)}{\rho(0)}]\times100\%$, at 10~K and 20~K, for several chosen magnetic field orientations ($\theta$ = 0$^\circ$, 15$^\circ$, 30$^\circ$, 45$^\circ$ and 60$^\circ$). The results are shown in Figs.~\ref{fig2}(c) and ~\ref{fig2}(d), respectively. For all magnetic field orientations, MR does not show any sign of saturation up to the highest magnetic field 9~T in our PPMS, and exhibits a similar field dependence H$^{n}$ (\emph{n} = 1.68$-$1.77) for the values of $\theta$ = 0$^\circ$, 30$^\circ$, 45$^\circ$ and 60$^\circ$. However, for $\theta$ = 15$^\circ$ we find \emph{n} = 1.90, a nearly quadratic scaling with the maximum MR of 9.43 $\times$ $10^3$$\%$ at 10~K, 9~T. We note that the field dependence of MR measured at 10~K and 20~K has the same power law for each field orientation, indicating that MR can be described by the Kohler scaling law \cite{pippard1989magnetoresistance}

\begin{eqnarray}
\textit{MR} = \frac{\Delta\rho_{xx}(T,H)}{\rho_{0}(T)} = \alpha(\textit{H}/\rho_{0})^{m}.
\end{eqnarray}

\begin{figure}[!htbp]
\centering
\includegraphics[width=8.6cm]{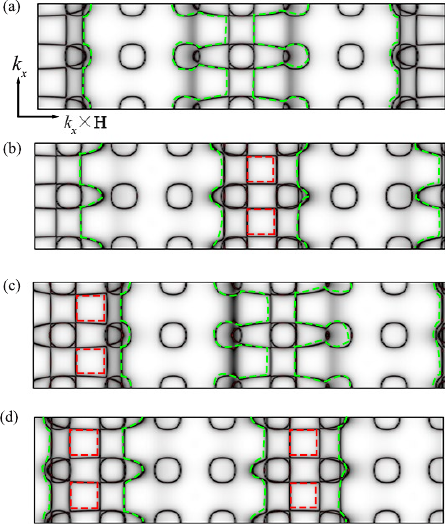}
\caption{Typical cross-sections of the FS of ReO$_3$  projected onto the \emph{k}$_x$ $\times$ \emph{H} plane, for the magnetic field orientation characterized by $\theta$ = 15$^\circ$. The horizontal axis corresponding to the \emph{k}$_x$ $\times$ \emph{H} direction, while the vertical to the \emph{k}$_x$ direction. The plane in panel (a) passes through the $\Gamma$ point, while the planes in panels (b), (c) and (d) pass through points (0,  0.1$\pi$/a, 0.1$\pi$/a), (0, 0.5$\pi$/a, 0.5$\pi$/a) and (0, 0.9$\pi$/a, 0.9$\pi$/a), respectively. The red dashed lines highlight the closed hole orbits while the green dashed lines indicate open electron orbits along the \emph{k}$_x$ direction.}\label{fig3}
\end{figure}

\begin{figure}[!htbp]
\centering
\includegraphics[width=8.6cm]{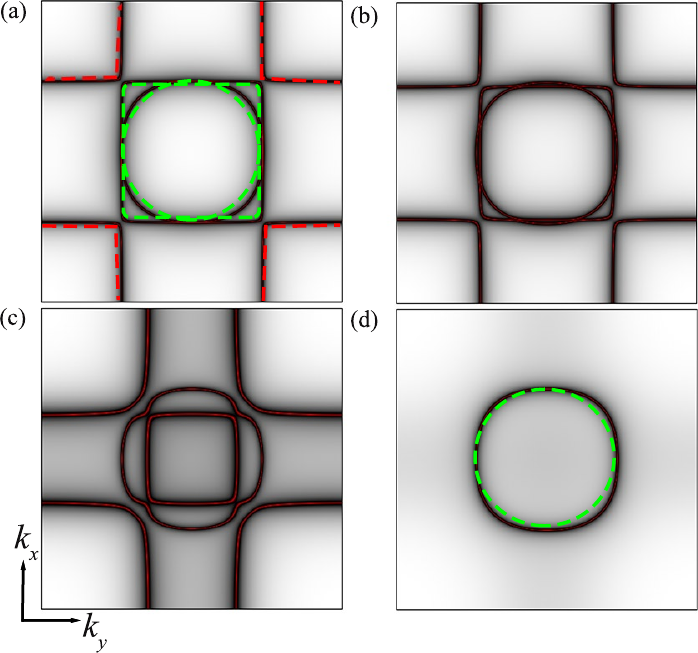}
\caption{Typical cross-sections of the FS of ReO$_3$ in the \emph{k}$_x$-\emph{k}$_y$ plane corresponding to (a) \emph{ k}$_z$ = 0, (b) \emph{k}$_z$ = 0.1$\pi$/a, (c) \emph{k}$_z$ = 0.2$\pi$/a, (d) \emph{k}$_z$ = 0.3$\pi$/a, for the \emph{H} $\parallel \emph{c}$ axis ($\theta$ = 0$^{\circ}$) orientation. Green and red dashed lines show the closed electron and hole orbits, respectively.}\label{fig4}
\end{figure}

In order to understand the quadratic magnetic field dependence for the $\theta$ = 15$^{\circ}$ orientation, we plot the representative orbits perpendicular to the magnetic field in Figs.~\ref{fig3}(a)$-$ \ref{fig3}(d), as well as in Fig.  \ref{fig1}(h). The red dashed lines highlight the closed hole orbits, while the green dashed lines indicate the open electron orbits along the $k_{x}$ direction. Here, the square-shaped red orbits originate from joining the electron pocket fragments in the adjacent periodic replicas of the Brillouin zone (BZ). These are the hole orbits rather than electron orbits. Thus, the non-saturating MR with a quadratic magnetic field dependence ($B^{1.9}$) for the $\theta$ = 15$^{\circ}$ orientation originates from the existence of open orbits. On the other hand, for the $\theta$ = 0$^{\circ}$ (\emph{H} $\parallel \emph{c}$ axis) orientation, representative orbits perpendicular to the magnetic field are shown in Figs.~\ref{fig4}(a)$-$\ref{fig4}(d). The green and red dashed lines indicate the closed electron and hole orbits, respectively, in which the square-shaped red orbits originate from joining the electron pocket ($\gamma$) fragments in the adjacent periodic replicas of BZ, \emph{i}.\emph{e}. the $\gamma_{2}$ orbit in Fig.~\ref{fig1}(g). In this case,  complete compensation of the two kinds of charge carriers can be achieved and confirmed by the Hall resistivity measurements discussed below.

\begin{figure}[!htbp]
\centering
\includegraphics[width=8.6cm]{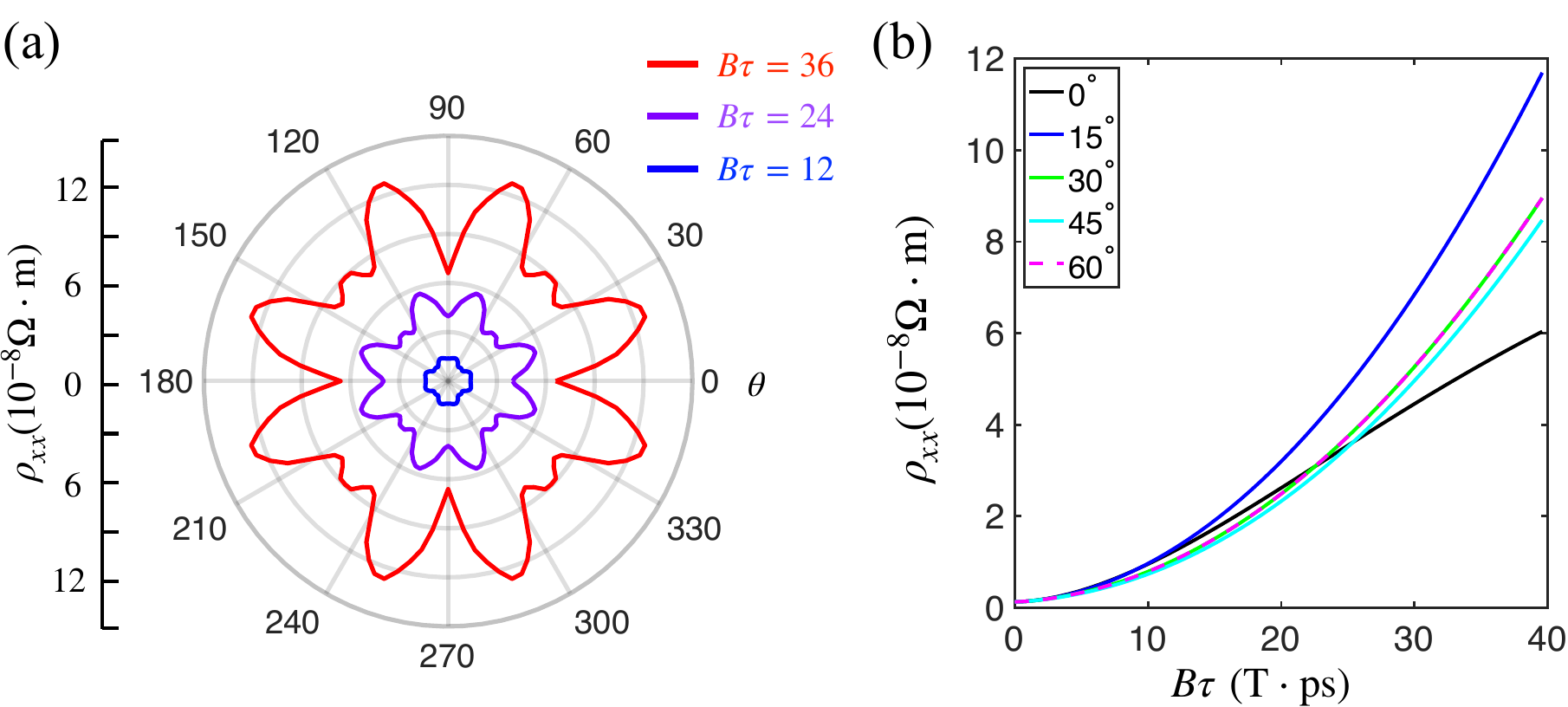}
\caption{(a) Calculated anisotropy of resistivity $\rho_{xx}$ for magnetic
field rotated in the $b$-$c$ plane agrees well with experiment results in
Fig. 2(b). (b) MR as a function of the magnitude
of magnetic field for the five different directions indicated by the $\theta$ values. \label{fig5}}
\end{figure}

Figure~\ref{fig5} shows the results of our numerical simulations for the
resistivity anisotropy and the magnetic field dependence of
MR. Figure~\ref{fig5}(a) shows calculated anisotropy of resistivity for $H$ rotated in the
$b$-$c$ plane, which agrees well with our measurements shown in Fig.~\ref{fig2}(b). The angular dependence of MR shows fourfold symmetry caused by the symmetry of crystal structure. The calculated magnetic field dependence of MR
also exhibits a sub-quadratic behavior, \emph{i}.\emph{e}., MR scales as $H^{1.9}$ for $\theta = 15^{\circ}$. All
calculated MR results for ReO$_3$ are well consistent with the experimental results
discussed above, which indicates that the topology of FS plays the crucial role in defining MR in material.

\begin{figure}[!htbp]
\centering
\includegraphics[width=8.6cm]{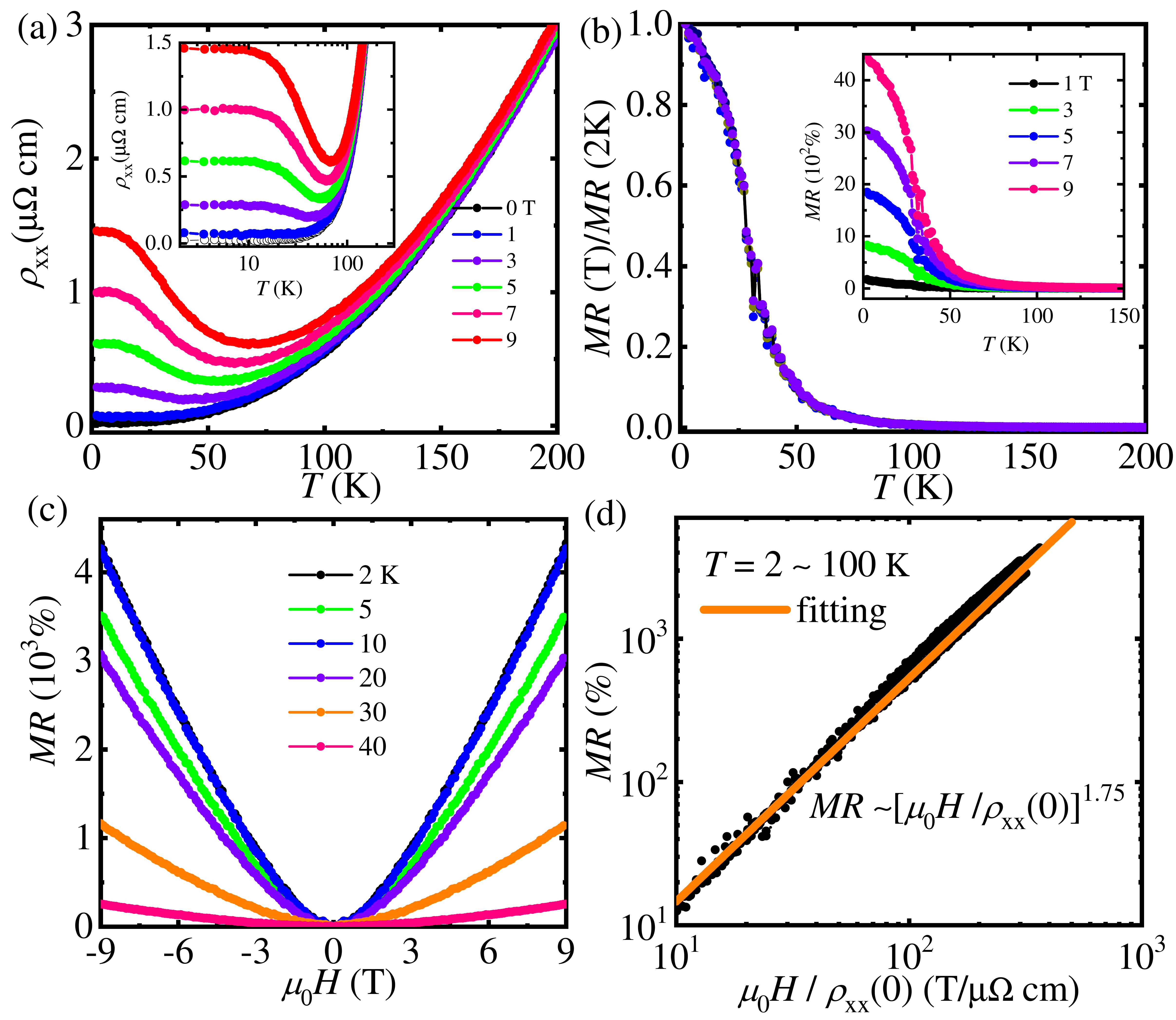}
\caption{(a) Temperature dependence of longitudinal resistivity $\rho$$_{xx}$(\emph{T})
measured at different magnetic fields
for the \emph{H} $\parallel \emph{c}$ orientation. (b) Temperature dependence of the MR normalized by its value at 2~K at various magnetic fields. The inset is the MR data as a function of temperature. (c) Field dependence of MR of ReO$_3$ at various temperatures. (d) MR as a function of \emph{H}/$\rho$$_{xx}$(0) plotted on a log scale.}\label{fig6}
\end{figure}

Figure~\ref{fig6} summarizes resistivity $\rho_{xx}$(\emph{T}, \emph{H}) measured at various temperatures and different magnetic fields with \emph{I} $\parallel \emph{a}$ axis, \emph{H} $\perp$ (001) plane ($\theta$ = 0$^{\circ}$) in ReO$_{3}$. The measured resistivity is remarkably enhanced by magnetic field at lower temperatures, and the field-induced up-turn is observed. The normalized MR has the same temperature dependence at various fields [see Fig.~\ref{fig6}(b)]. Figure~\ref{fig6}(c) displays MR as a function of magnetic field at various temperatures, which reaches 4.33 $\times$ 10$^{3}\%$ at 2~K and 9~T, and does not show any sign of saturation. The MR can be described by the Kohler scaling law [see Fig.~\ref{fig6}(d)] with fitting parameters $\alpha$ = 0.34 ($\mu\Omega$ cm/T)$^{1.75}$ and \emph{m} = 1.75. As an ``ordinary" nonmagnetic metal, ReO$_{3}$ exhibits all the common behaviors observed in many trivial or nontrivial topological semimetals \cite{takatsu2013extremely,mun2012magnetic,yuan2016large,huang2015observation,ali2014large,chen2016extremely} with XMR, which seems to be unexpected.

\begin{figure}[!htbp]
\centering
\includegraphics[width=8.6cm]{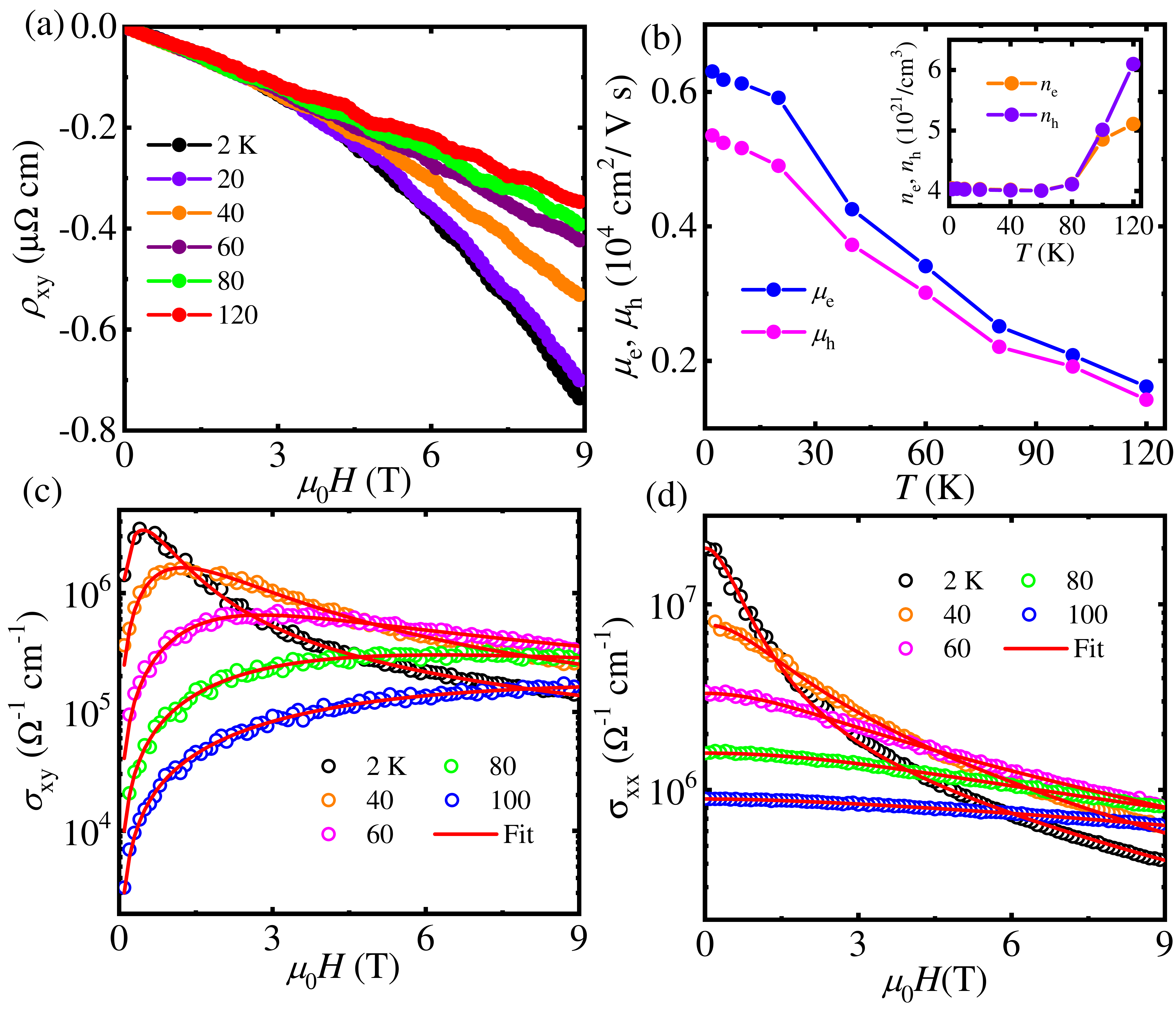}
\caption{(a) Field dependence of Hall resistivity $\rho_{xy}$ measured for \emph{H} $\parallel \emph{c}$ axis at different temperatures. (b) Charge-carrier mobilities, $\mu_e$ and $\mu_h$, and (inset) carrier concentrations, \emph{n}$_e$ and \emph{n}$_h$, as a function of temperature extracted from the two-carrier model. Components of the conductivity tensor, $\sigma_{xy}$ and $\sigma_{xx}$, shown in panels (c) and (d), respectively, as functions of magnetic field for temperatures ranging from 2 to 100~K. Dots represent experimental data and red solid lines the fitting curves based on the two-carrier model.}\label{fig7}
\end{figure}

\begin{figure}[!htbp]
\centering
\includegraphics[width=8.6cm]{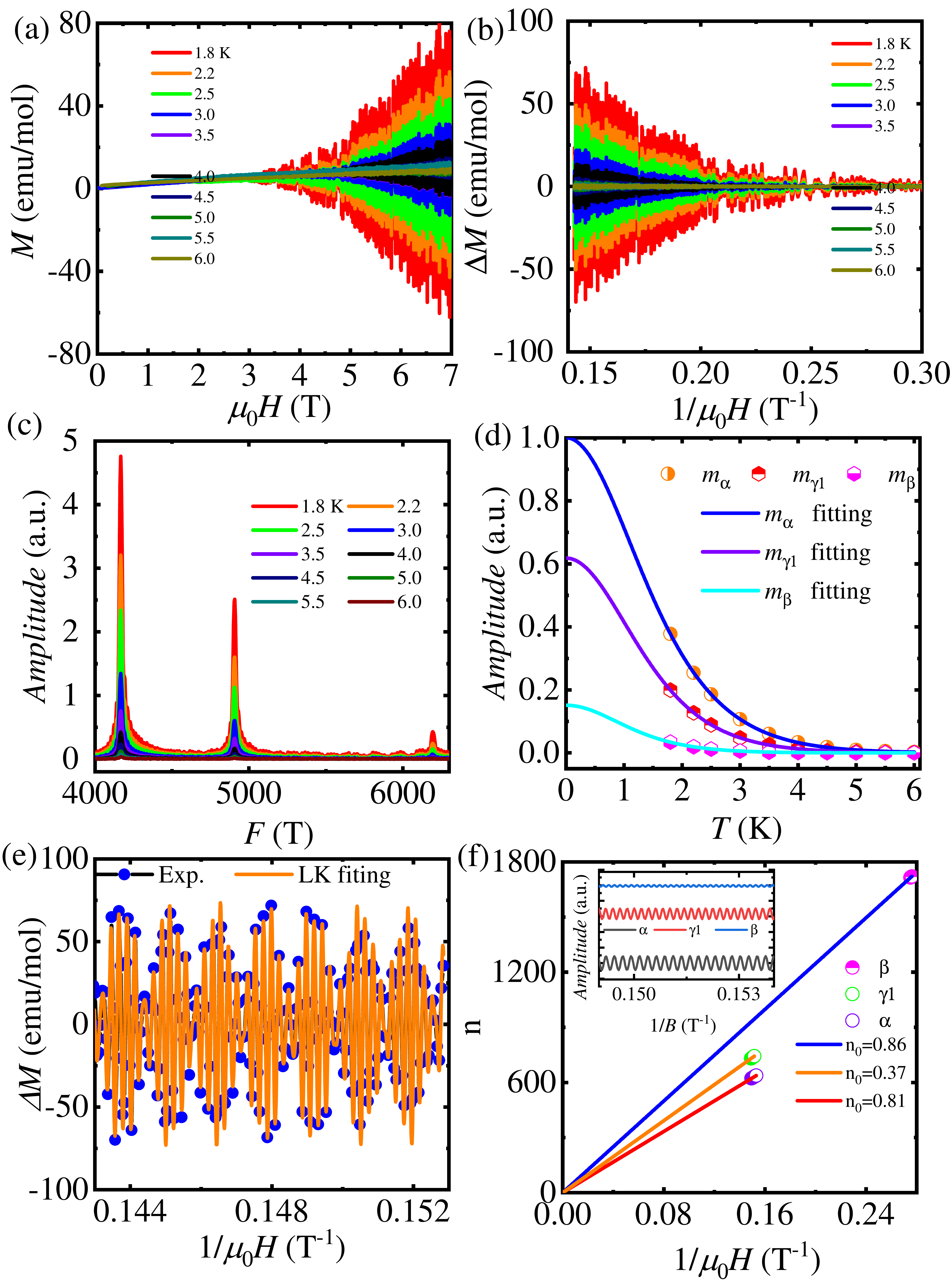}
\caption{(a) Magnetization as a function of field measured at various temperatures. (b) The amplitude of dHvA oscillations plotted as a function of 1/$\mu_0$\emph{H}. (c) Fourier transform (FT) spectra of the dHvA oscillations measured between 1.8~K and 6.0~K. (d) Temperature dependence of relative FT amplitudes for each frequency and the fitting results by \emph{R}$_T$. (e) The fitting of dHvA oscillations at 1.8~K by the multi-band LK formula. (f) Landau-level indices fan diagram for the three filtered frequencies plotted as functions of 1/$\mu_0$\emph{H}, respectively, and the filtered waves of the three frequencies (inset).}\label{fig8}
\end{figure}

In fact, as discussed above, for this particular magnetic field orientation ($\theta$ = 0$^{\circ}$), there are indeed two kinds of charge carriers in ReO$_{3}$, evidenced by the nonlinear field dependence of Hall resistivity measured at various temperatures [see Fig.~\ref{fig7}(a)]. Following the analysis of $\gamma$-MoTe$_2$ by Zhou \emph{et al}. \cite{zhou2016hall}, as well as our work on MoO$_2$ \cite{PhysRevB.102.165133}, we analysed the longitudinal and Hall resistivity data by using the semiclassical two-carrier model. In this model, the conductivity tensor, in its complex representation, which is given by \cite{ali2014large}
\begin{eqnarray}
  \sigma = \frac{en_{e}\mu_{e}}{1+i\mu_{e}\mu_0H}+\frac{en_{h}\mu_{h}}{1-i\mu_{h}\mu_0H} ,
\end{eqnarray}
where \emph{n}$_e$ and \emph{n}$_h$ denote the carrier concentrations, $\mu_e$ and $\mu_h$ denote the mobilities of electrons and holes, respectively. To evaluate the carrier densities and their mobilities, we calculated the Hall conductivity $\sigma_{xy}$ = $-$$ \rho_{xy}/(\rho_{xx}^{2}+\rho_{xy}^{2})$, and the longitudinal conductivity $\sigma_{xx}$ = $\rho_{xx}/(\rho_{xx}^{2}+\rho_{xy}^{2})$ by using the original experimental $\rho_{xy}$(\emph{H}) and $\rho_{xx}$(\emph{H}) data. Then, we fit both $\sigma_{xy}$(\emph{H}) and $\sigma_{xx}$(\emph{H}) data by using the same fitting parameters and their field dependences given by \cite{zhou2016hall}
\begin{eqnarray}
\sigma_{xy} =  \frac{e\mu_0Hn_{h}\mu_{h}^{2}}{1+\mu_{h}^{2}\mu_0^{2}H^{2}}-\frac{e\mu_0Hn_{e}\mu_{e}^{2}}{1+\mu_{e}^{2}\mu_0^{2}H^{2}},
\end{eqnarray}
\begin{eqnarray}
\sigma_{xx} = \frac{en_{h}\mu_{h}}{1+\mu_{h}^{2}\mu_0^{2}H^{2}}+\frac{en_{e}\mu_{e}}{1+\mu_{e}^{2}\mu_0^{2}H^{2}}.
\end{eqnarray}

Figures~\ref{fig7}(c) and~\ref{fig7}(d) display the fits of both the $\sigma_{xy}$(\emph{H}) and $\sigma_{xx}$(\emph{H}) measured at \emph{T} = 2, 40, 60, 80 and 100~K, respectively. The excellent agreement between our experimental data and the two-carrier model over a broad range of temperatures confirms the coexistence of electrons and holes in ReO$_3$. Figure  \ref{fig7}(b) shows the \emph{ n}$_e$, \emph{n}$_h$, $\mu_e$ and $\mu_h$ values obtained by the fitting over the temperature range 2$-$120~K. It is remarkable that the \emph{n}$_e$ and \emph{n}$_h$ values are almost the same below 100~K, as shown in the inset of Fig.~\ref{fig7}(b), such as at 2~K, \emph{n}$_e$ = 4.04 $\times$ 10$^{21}$ cm$^{-3}$, and \emph{n}$_h$ = 4.03  $\times $ 10$^{21}$ cm$^{-3}$. These results indicate that the MR in ReO$_3$ metal for this particular magnetic field orientation (\emph{H} $\parallel$\emph{c}) results indeed from the perfect compensation of the two kinds of charge carriers, similar to that observed in many trivial and topologically nontrivial semimetals \cite{takatsu2013extremely,mun2012magnetic,yuan2016large,huang2015observation,ali2014large,chen2016extremely}. This verifies once again the leading role of the details of FS topology played in the magnetotransport.

Finally, in order to obtain additional information on the electronic structure, we measured the dHvA quantum oscillations in the isothermal magnetization,\emph{ M}(\emph{H}), for the \emph{H} $\parallel$ \emph{c} axis up to 7~T. As shown in Fig.~\ref{fig8}(a), clear dHvA oscillations in the \emph{M}(\emph{H}) curves were observed up to 6.0~K from 3.5~T. After subtracting a smooth background from the \emph{M}(\emph{H}) data at each temperature, periodic oscillations are visible in 1/\emph{H}, as shown in Fig.~\ref{fig8}(b). From the Fourier transform (FT) analysis, we derived three basic frequencies 4167~T ($F_\alpha$), 4908~T ($F_{\gamma 1}$) and 6194~T ($F_\beta$) [Fig.~\ref{fig8}(c)], which are consistent with the results reported by Schirber \emph{et al}. \cite{PhysRevB.24.692}. In general, the oscillatory magnetization of a three-dimensional (3D) system can be described by the Lifshitz-Kosevich (LK) formula \cite{lifshitz1956theory,shoenberg1984magnetic} with the Berry phase \cite{mikitik1999manifestation}
\begin{eqnarray}
\Delta M \varpropto -B^{1/2}R_{T}R_{D}R_{S} \sin[2\pi(\emph{F}/\emph{B}-\gamma-\delta)],
  \end{eqnarray}
 \begin{eqnarray}
  R_{T}=\alpha T\mu/B \sinh(\alpha T\mu/B),
  \end{eqnarray}
  \begin{eqnarray}
    R_{D}=\exp(-\alpha T_{D}\mu/B),
  \end{eqnarray}
  \begin{eqnarray}
  R_{S}=\cos(\pi g\mu/2).
   \end{eqnarray}
where $\mu$ is the ratio of effective cyclotron mass \emph{m}$^*$ to free electron mass \emph{m}$_0$. \emph{T}$_\emph{D}$ is the Dingle temperature, and  $\alpha$ = ($2\pi^2 $$\emph{k}_\emph{B} \emph{m}_0$)/($\hbar$e). The phase factor $\delta$ = 1/8 or $-$1/8 for three dimensional systems. The effective mass \emph{m}$^*$ can be obtained by fitting the temperature dependence of the oscillation amplitude \emph{R}$_\emph{T}$(T), as shown in Fig.~\ref{fig8}(d). For $F_\alpha = 4167$~T, $F_{\gamma 1} = 4908$~T and $F_\beta = 6194$~T, the obtained \emph{m}$^*$ are 0.42\emph{m}$_0$, 0.45\emph{m}$_0$ and 0.54\emph{m}$_0$, respectively, somewhat smaller than the calculated values by Schirber \emph{et al}. \cite{PhysRevB.24.692}. Using the fitted \emph{m}$^*$ as a known parameter, we can further fit the oscillation patterns at given temperatures [{\it e.g.} \emph{T} = 1.8~K, see Fig.~\ref{fig8}(e)] to the LK formula with three frequencies, from which quantum mobility and the Berry phase can be extracted. The fitted Dingle temperatures \emph{T}$_\emph{D}$ are 11.99~K, 11.48~K and 9.00~K, which corresponds to the quantum relaxation times $\tau_q$ = $\hbar$/(2$\pi$$\emph{$k_BT_D$}$) of 0.10~ps, 0.10~ps and 0.13~ps, respectively. The quantum mobilities $\mu_q$ (e$\tau$/\emph{m}$^*$) are 419 cm$^2$/V s, 391 cm$^2$/V~s, 423 cm$^2$/V~s for $F_\alpha$, $F_{\gamma 1}$  and $F_\beta$, respectively, as listed in Table I. The LK fit also yields a phase factor $-$ $\gamma$$-$ $\delta$ of $-$ 0.64 ($F_\alpha$), from which the Berry phase $\phi_\emph{B}$ is determined to be 1.97$\pi$ for $\delta$ = 1/8 and 1.4 $\pi$ for $\delta$ = $-$1/8. The phase factor for $F_{\gamma 1}$ is 0.38, with $\phi_\emph{B}$ of 0.01$\pi$ ($\delta$ = 1/8) and 1.51$\pi$ ($\delta$ = $-$1/8). Other results are displayed in Table I.

Similar Berry phase values can also be obtained from the commonly used Landau level fan diagram \cite{hu2017nearly} (\emph{i}.\emph{e}. the LL index \emph{n} as a function of the inverse of magnetic field 1/$\emph{B}_n$). According to customary practice, the integer LL indices \emph{n} should be assigned when the Fermi level lies between two adjacent LLs \cite{xiong2012high}, where the density of states (DOS) near the Fermi level (\emph{E}$_F$) reaches a minimum. Given that the oscillatory magnetic susceptibility is proportional to the oscillatory DOS ({$\emph{E}_F$}) [\emph{i}.\emph{e}. $\Delta$(d\emph{M}/d\emph{B}) $\propto$ $\Delta$DOS($\emph{E}_F$)] and that the minima of $\Delta$\emph{M} and d($\Delta$\emph{M})/d\emph{B} are shifted by $\pi$/2, the minima of $\Delta$\emph{M} should be assigned to \emph{n}$-$1/4. The established LL fan diagram based on this definition is shown in Fig.~\ref{fig8}(f). The extrapolation of the linear fit in the fan diagram yields an intercept $\emph{n}_0$ = 0.81, which appears to correspond to a Berry phase of $\phi _\emph{B}$ = 2$\pi$(0.81+$\delta$), that is  1.87$\pi$ ($\delta$ = 1/8) and 1.35$\pi$ ($\delta$ = $-$ 1/8) for the $F_\alpha$ band. These results are consistent with the results of the LK formula. Besides, we also obtained the Berry phase for the $F_{\gamma 1}$ band and $F_\beta$ band  as listed in Table I. It is well known that topologically non-trivial materials requires a non-trivial $\pi$ Berry phase, while for the trivial materials, the Berry phase equals 0 or 2$\pi$ . In our sample, the Berry phase is away from the $\pi$, hence we conclude ReO$_3$ is a topologically trivial material.

\begin{table}[htbp]
  \renewcommand\arraystretch{1.2}
  \centering
  \caption{The parameters obtained by fitting the dHvA data for ReO$_3$.}\label{1}
  \setlength{\tabcolsep}{4.0mm}
  {
  \begin{center}
  \begin{tabular}{ccccc}
    \hline
    \hline
    Parameters &F$_{\alpha}$(LK)&F$_{\gamma1}$(LK)&F$_{\beta}$(LK) \\

    \hline
    Frequency (T)& 4167 & 4908 & 6194  \\
    \textit{m}$^{*}$/\textit{m}$_{0}$ & 0.42 & 0.45& 0.54  \\
    \emph{T}$_{D}$ (K)& 11.99 & 11.48&9.00 \\
    $\tau_{q}$ (ps)& 0.10 &0.10 &0.13 \\
    $\mu_{q}$ (cm$^{2}$/Vs)& 419 & 391&423 \\
    $\varphi_{B}$($\delta$=+1/8) &1.97$\pi$&0.01$\pi$&0.33$\pi$\\
    $\varphi_{B}$($\delta$=-1/8) &1.47$\pi$&1.51$\pi$&1.83$\pi$\\
  Parameters &F$_{\alpha}$(LL)& F$_{\gamma1}$(LL) &F$_{\beta}$(LL) \\
 $\varphi_{B}$($\delta$=+1/8) &1.87$\pi$&0.97$\pi$&-0.03$\pi$\\
 $\varphi_{B}$($\delta$=-1/8) &1.35$\pi$&0.29$\pi$&1.47$\pi$\\
    \botrule

  \end{tabular}
   \end{center}
  }
\end{table}

\section{IV. SUMMARY}

In summary, for successfully synthesized ReO$_3$ crystals, we measured $\rho_{xx}$(\emph{T}, \emph{H}), Hall resistivity, $\rho_{xy}$(\emph{T}, \emph{H}), and dHvA oscillations, as well as calculated the electronic band structure and FS to study the anisotropy of MR. It was found that for magnetic field applied along the \emph{c} axis, the MR exhibits a non-saturating \emph{H}$^{1.75}$ dependence, which arises from the carrier compensation as supported by the $\rho_{xy}$(\emph{T}, \emph{H}) measurements. For \emph{H} oriented along other directions, a similar non-saturating  \emph{H}$^{n}$ (\emph{n} = 1.68$-$1.90) dependence of MR was observed, but in this case it stems from the existence of open orbits extending along the $k_{x}$ direction. As an ``ordinary" metal, ReO$_3$ exhibits all the characteristics of XMR semimetals, which is attributed to its peculiar FS, implying the details of FS topology being the key factor underlying the observed XMR in materials.

\section{\textbf{ACKNOWLEDGMENTS}}

This research is supported by the National Key R$\&$D Program of China under Grants No. 2016YFA0300402, and the National Natural Science Foundation of China ( Grants No. NSFC-12074335 and 11974095), the Zhejiang Natural Science Foundation (No. LY16A040012) and the Fundamental Research Funds for the Central Universities. S.N.Z, Q.S.W. and O.V.Y. acknowledge support by the NCCR Marvel. First-principles calculations were partly performed at the Swiss National Supercomputing Centre (CSCS) under projects s1008 and mr27 and the facilities of Scientific IT and Application Support Center of EPFL.
\bibliography{document}
\end{document}